\documentclass[fleqn,twoside]{article}
\usepackage{espcrc2}
\usepackage{psfig}

\def\D{{\rm d}}
\def\I{{\rm i}}

\begin{document}

\title{
{\vspace{-1.2em} \parbox{\hsize}{\hbox to \hsize 
{\hss  \normalsize MS-TP-01-7, IFUP-TH 2001/31}}} \\
Hamiltonian Monte Carlo simulation \\
of the two-dimensional Wess-Zumino model}

\author{
Matteo Beccaria\address{Dipartimento di Fisica
dell'Universit\`a di Lecce and I.N.F.N., Sezione di Lecce},
Massimo Campostrini\address{Dipartimento di Fisica
dell'Universit\`a di Pisa and I.N.F.N., Sezione di Pisa}%
\thanks{Presented by M.\ Campostrini} and
Alessandra Feo\address{Institut f\"ur Theoretische Physik,
Westf\"alische Wilhelms-Universit\"at, M\"unster}
}

\begin{abstract}
We study a Hamiltonian lattice version of the two-dimensional
Wess-Zumino model. Preliminary results obtained by Quantum
Monte Carlo with a many-parameter guiding wave function are presented.
We analyze the pattern of supersymmetry breaking by measuring 
the ground state energy and a set of supersymmetric Ward identities. 
The algorithm is quite effective and allows very precise measurements.
\end{abstract}

\maketitle

\section{INTRODUCTION}

Numerical simulations of lattice field theories are usually performed
in the Lagrangian formulation.  The alternative approach of the
Hamiltonian formulation \cite{KS} has received very little attention
so far.  Nonetheless, we think there are very good reasons to
develop numerical simulation techniques for the Hamiltonian approach:
powerful many-body techniques are available \cite{QMC}, which allow
the direct computation of the vacuum wave function properties;
fermions are implemented directly and need not be integrated out;
property like the mass spectrum are more immediate.  A check of
universality between the Lagrangian and the Hamiltonian formalism
would also be very welcome.

\section{THE MODEL}
The continuum two-dimensional Wess-Zumino model is defined by the
supersymmetry charges
\begin{eqnarray*}
Q_{1,2} &=& \int \D x \biggl[p(x)\,\psi^{1,2}(x) \\
    &&\qquad\qquad-\biggl({\partial\phi\over\partial x}
    \pm V(\phi(x))\biggr)\psi^{2,1}(x) \biggr],
\end{eqnarray*}
where $\phi(x)$ is a real scalar field, $p(x)$ is its conjugate
momentum, and $\psi(x)$ is a Majorana fermion.  The charges obey the
$N=1$ supersymmetry algebra
\[
Q_1^2 = Q_2^2 = P^0 \equiv H, \qquad \{Q_1,Q_2\} = 2 P^1 \equiv 2 P,
\]
which we have written in the Majorana basis $\gamma_0 = C = -\sigma_2$,
$\gamma_1 = \I\sigma_3$. Since $P$ is not conserved on the lattice, a
lattice formulation of a supersymmetric model must break the algebra
explicitly.  A very important advantage of the Hamiltonian formulation
is that, since $H$ is conserved exactly, we can maintain a
1-dimensional supersymmetry subalgebra, e.g., $Q_1^2 = H$, and expect
to recover the rest in the continuum limit, together with Lorentz
invariance \cite{ERS}; this subalgebra is enough to guarantee some of
the most important property of supersymmetry, including pairing of
fermionic and bosonic states of nonzero energy; spontaneous breaking
of supersymmetry is equivalent to nonzero ground-state energy $E_0$.

We adopt the lattice formulation \cite{RS}
\begin{eqnarray*}
Q \equiv Q_1 &=& \sum_{n=1}^L \biggl[ p_n \psi^1_n \\
    &&\qquad-\, \biggl({\phi_{n+1} - \phi_{n-1}\over2}
        + V(\phi_n)\biggr)\psi^2_n \biggr],
\end{eqnarray*}
with canonical (anti)commutation rules, and $H=Q^2$.  We can replace
the Majorana fields $\psi^{1,2}$ by one Dirac field $\chi$ by the
clever transformation \cite{RS}
\[
\psi^{1,2}_n = 
\frac{(-1)^n \mp \I}{2\I^n}(\chi_n^\dagger \pm \I\chi_n),
\]
obtaining the Hamiltonian
\begin{eqnarray*}
&&H = {1\over2} \sum_{n=1}^L \biggl[ p_n^2 + 
\left({\phi_{n+1} - \phi_{n-1}\over2} + V(\phi_n)\right)^{\!2} \\
&& -\, (\chi^\dagger_n\chi_{n+1} + h.c.) + (-1)^n V'(\phi_n)
\left(2\chi^\dagger_n\chi_n-1\right) \biggr]
\end{eqnarray*}
which is explicitly free of sign problems, except potentially at the
boundary; we adopt {\em free boundary conditions} to avoid the sign
problem altogether \cite{Trento}.

In strong coupling, the model reduces to a supersymmetric quantum
mechanics for each site; supersymmetry is broken if and only if the
degree of the prepotential $V$ is even \cite{Witten}.
In the continuum (and on the lattice in weak coupling), supersymmetry
is broken at tree level if and only if $V$ has no zeroes.
The predictions of strong coupling and weak coupling are quite
different, and it is interesting to study both numerically and
analytically the crossover from strong to weak coupling.

As a benchmark of the case where supersymmetry is unbroken, we consider the 
cubic prepotential $V=\phi^3$. In the more interesting case of broken
supersymmetry we study the quadratic prepotential $V=\phi^2+\lambda_0$.
Here, spontaneous symmetry breaking is expected to occur at $\lambda_0>0$, 
while for $\lambda_0<0$ the scenario is less clear; in fact, weak coupling
perturbation theory predicts a symmetric ground state while strong coupling 
predicts broken supersymmetry with $E_0\sim\exp(c\lambda_0)$ for $\lambda_0\to-\infty$.

\section{MONTE CARLO SIMULATIONS}

We perform our simulations by the Green Function Monte Carlo algorithm
\cite{QMC} with the guiding wave function associated to the trial
ground state
\begin{eqnarray*}
|\Psi_0\rangle_{\rm trial} &=& 
\exp\Bigl[\textstyle\sum_n F_n\Bigr] 
|\Psi_0\rangle_{\rm free}, \\
F_n &=& P_1(\phi_n)+(-1)^n(\chi_n^\dagger\chi_n-1/2)P_2(\phi_n),
\end{eqnarray*}
where $|\Psi_0\rangle_{\rm free}$ is the free ($V=0$) ground state and
$P_{1,2}$ are polynomials optimized according to the algorithm
described in Ref.\ \cite{MB}; see Ref.~\cite{Trento} for more details.
We wish to remark that, in order to keep the variance of observables
finite as the simulation proceeds, it is necessary to simulate a
population of $K$ {\em walkers} (field configurations at fixed time),
and extrapolate the results to $K\to\infty$.

The trial wave function is an important feature of our simulations,
and we are able to determine accurately the free parameters.  The trial
wave function allows a substantially improvement of the quality of our
numerical results; moreover, since it is an approximation to the
exact ground state, it contains important physical information about
the model.

This algorithm can be parallelized effectively on a network of PCs
connected through Ethernet; our MPI code reaches 90\% efficiency.

We measure the ground-state energy $E_0$, which is the simplest and
most precise observable to measure in a Quantum Monte Carlo, and a
number of observables of the form $\{Q,X_q\}$,
$X_q=\sum_n\phi^q_n\psi^2_n$, which are zero for unbroken
supersymmetry (for any fermionic operator $X$,
$\langle\{Q,X\}\rangle=0$ is a supersymmetry Ward identity).

\begin{figure}[tb]
\null\vskip2mm
\centerline{\psfig{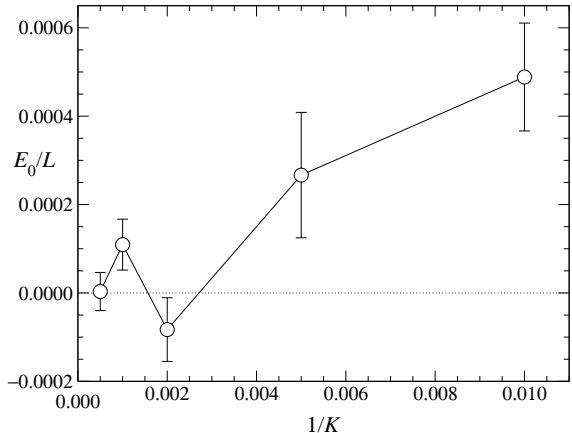}}
\vskip-8mm
\caption{$E_0/L$ vs.\ $1/K$ for $V = \phi^3$.}
\label{fig:E0-cubic}
\end{figure}

Fig.\ \ref{fig:E0-cubic} shows the ground-state energy per site as a
function of $1/K$ for the cubic prepotential $V = \phi^3$ and $L=10$.
The evidence for unbroken supersymmetry is quite convincing.  The
bosonic and fermionic contribution to $E_0/L$ are $\simeq \pm0.7$: we
are observing a cancellation of the order of $10^{-4}$.  Similar
conclusions can be drawn by looking at $\{Q,X_q\}$, with smaller
numerical accuracy.

\begin{figure}[tb]
\null\vskip2mm
\centerline{\psfig{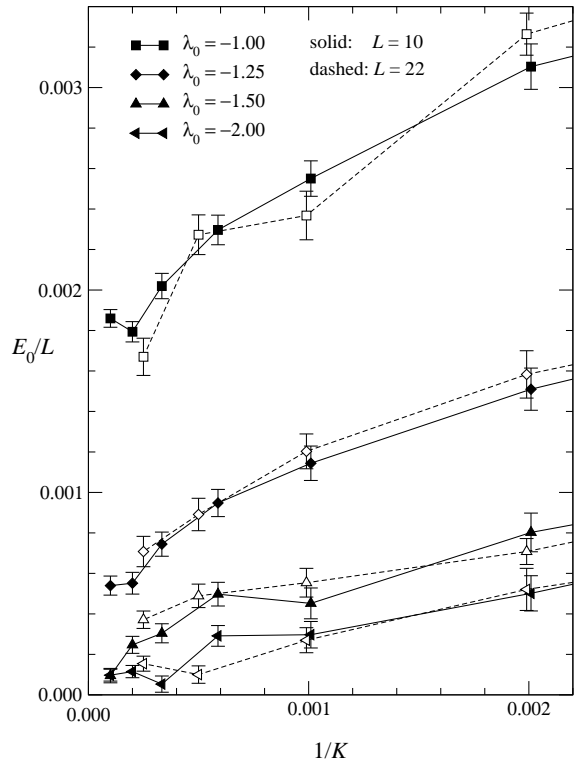}}
\vskip-8mm
\caption{$E_0/L$ vs.\ $1/K$ for $V = \phi^2 + \lambda_0$.}
\label{fig:E0-quadratic}
\end{figure}

Fig.\ \ref{fig:E0-quadratic} shows the ground-state energy per site as
a function of $1/K$ for the quadratic prepotential $V = \phi^2 +
\lambda_0$, for $\lambda_0=-1$, $-1.25$, $-1.5$, and $-2$ and $L=10$,
$22$.  The data are insensitive to the value of $L$, and they are
consistent with a linear extrapolation to $1/K\to0$.  Supersymmetry
breaking is clear for $\lambda_0\ge-1.25$; for smaller values of
$\lambda_0$, we cannot discriminate between $E_0=0$ and an
exponentially decreasing $E_0$.  As in the cubic case, the study of
$\{Q,X_q\}$ gives similar conclusions, with smaller numerical
accuracy.

\section{CONCLUSIONS}

The Hamiltonian approach is a worthwhile alternative to the usual
Lagrangian approach to lattice field theory.  Accurate Monte Carlo
simulations can be performed and compared to analytical predictions,
e.g., strong coupling computations.  Fermions can be implemented
directly and need not be integrated out, thus avoiding a non-local
determinant; for certain models in $1+1$ dimensions, like the one
studied here, they present no additional costs compared to bosons.  It
should be observed however, that in higher dimension fermions lead to
{\em sign problems} which need to be solved \cite{Sign}.  Another
important advantage of the Hamiltonian formalism is the possibility of
preserving exactly a 1-dimensional supersymmetry algebra.

In the light of the encouraging preliminary results presented here, we
plan to extend our study in several directions.  We are computing the
strong-coupling expansion, and we will study the weak-coupling
expansion and the perturbative continuum limit.  We will investigate
numerically the transition from strong to weak coupling, looking at
the pattern of supersymmetry breaking.

\begin{thebibliography}{9}

\bibitem{KS} J.~Kogut, L.~I.~Susskind, 
Phys.\ Rev.\ {\bf D11} (1975) 395; 
J.~Kogut, Rev.\ Mod.\ Phys.\ {\bf 51} (1979) 659.

\bibitem{QMC} W.~von der Linden,
Phys.\ Rept.\ {\bf 220} (1992) 53.

\bibitem{ERS} S.~Elitzur, E.~Rabinovici, A.~Schwimmer, 
Phys.\ Lett.\ {\bf B119} (1982) 165. 

\bibitem{RS} J.~Ranft, A.~Schiller, 
Phys.\ Lett.\ {\bf B138} (1984) 166.

\bibitem{Trento} M.~Beccaria, M.~Campostrini, A.~Feo, 
hep-lat/0109005, to be published in
Proc.\ of the Quantum Monte Carlo meeting,
Trento, Italy, July 3-6, 2001.

\bibitem{Witten} E.~Witten, Nucl.\ Phys.\ {\bf B188} (1981) 513;
Nucl.\ Phys.\ {\bf B202} (1982) 253.

\bibitem{MB} M.~Beccaria, 
Phys.\ Rev.\ {\bf D61} (2000) 114503.

\bibitem{Sign} E.~Y.~Loh {\em et al.},
Phys.\ Rev.\ {\bf B41} (1990) 9301;
S.~Chandrasekharan, U.-J.~Wiese,
Phys.\ Rev.\ Lett.\ {\bf 83} (1999) 3116.

\end{thebibliography}
\end{document}